\documentclass[]{spie}  %>>> use for US letter paper

\renewcommand{\baselinestretch}{1.65} % Change to 1.65 for double spacing
\usepackage{subcaption}
\usepackage{amsmath,amsfonts,amssymb}
\usepackage{graphicx}
\usepackage[colorlinks=true, allcolors=blue]{hyperref}
\usepackage[leftcaption]{sidecap}

\title{Re-defining Radiology Quality Assurance (QA) - Artificial Intelligence 
(AI)-Based QA by Restricted Investigation of Unequal Scores 
(AQUARIUS) }

\author[a,b,c,d]{Axel Wismüller}
\author[a]{Larry Stockmaster}
\author[b]{M. Ali Vosoughi}

\affil[a]{Department of Imaging Sciences, University of Rochester, NY, USA}
\affil[b]{Department of Electrical and Computer Engineering, University of Rochester, NY, USA}
\affil[c]{Department of Biomedical Engineering, University of Rochester, NY, USA}
\affil[d]{Faculty of Medicine and Institute of Clinical Radiology, Ludwig Maximilian University,
Munich, Germany}

\authorinfo{Further author information: (Send correspondence to Axel Wismüller)\\Axel Wismüller: E-mail: axel.wismueller@gmail.com}

\begin{document} 
\maketitle

\begin{abstract}
There is an urgent need for streamlining radiology Quality Assurance (QA) programs to make them better and faster. Here, we present a novel approach, \textbf{A}rtificial Intelligence (AI)-Based \textbf{Qu}ality \textbf{A}ssurance by \textbf{R}estricted \textbf{I}nvestigation of \textbf{U}nequal \textbf{S}cores (AQUARIUS), for re-defining radiology QA, which reduces human effort by up to several orders of magnitude over existing approaches. AQUARIUS typically includes automatic comparison of AI-based image analysis with natural language processing (NLP) on radiology reports. Only the usually small subset of cases with discordant reads is subsequently reviewed by human experts. To demonstrate the clinical applicability of AQUARIUS, we performed a clinical QA study on Intracranial Hemorrhage (ICH) detection in 1936 head CT scans from a large academic hospital. Immediately following image acquisition, scans were automatically analyzed for ICH using a commercially available software (Aidoc, Tel Aviv, Israel). Cases rated positive for ICH by AI (ICH-AI+) were automatically flagged in radiologists' reading worklists, where flagging was randomly switched off with probability 50\%. Using AQUARIUS with NLP on final radiology reports and targeted expert neuroradiology review of only 29 discordantly classified cases reduced the human QA effort by 98.5\%, where we found a total of six non-reported true ICH+ cases, with radiologists' missed ICH detection rates of 0.52\% and 2.5\% for flagged and non-flagged cases, respectively. We conclude that AQUARIUS, by combining AI-based image analysis with NLP-based pre-selection of cases for targeted human expert review, can efficiently identify missed findings in radiology studies and significantly expedite radiology QA programs in a hybrid human-machine interoperability approach. 
\end{abstract}

% Include a list of keywords after the abstract 

\section{MOTIVATION} \label{sec:intro}  
The provision of high-quality health care is the goal of all medical services. In the case of radiology quality assurance (QA) programs, patient selection, the conduct of the examination, and the interpretation of the results can all have an impact on the achievement of this goal, such as by improving the diagnostic information content, reducing radiation exposure, reducing medical costs, and improving departmental management [\citeonline{1_SPIE22_davies1983quality}]. QA programs thus contribute to the provision of high-quality health care. 
 
Specifically, QA programs in radiology are of immense value in improving patient safety, optimizing the clinical usefulness of medical imaging exams, and increasing financial profits of radiology practices by lowering operational costs. Yet, contemporary approaches to radiology QA, such as widely used methods of randomly selecting cases for human review, suffer from limited effectiveness and efficiency. Hence, there is an urgent need for streamlining radiology QA programs to make them better and faster. Here, we present a novel hybrid human-machine interoperability approach, \textbf{A}rtificial Intelligence (AI)-Based \textbf{Qu}ality \textbf{A}ssurance by \textbf{R}estricted \textbf{I}nvestigation of \textbf{U}nequal \textbf{S}cores (AQUARIUS), for re-defining radiology QA, which has the potential of reducing human QA effort by several orders of magnitude over existing approaches.  
 
In the following, we will explain the basic idea of AQUARIUS, describe its clinical application within a retrospective clinical trial for QA in AI-based detection of Intracranial Hemorrhage (ICH) in emergent care head CT scans, and will report initial radiology QA results for this clinical application.  
\newline
This work is embedded in our group’s endeavor to expedite artificial intelligence in biomedical imaging by means of advanced pattern recognition and machine learning methods for computational radiology and radiomics, e.g., [\citeonline{
12_wismuller2006exploratory,
13_wismuller1998neural,
14_wismuller2002deformable,
15_behrends2003segmentation,
16_wismuller1997neural,
17_bunte2010exploratory,
18_wismuller1998deformable,
19_wismuller2009exploration,
20_wismuller2009method,
22_huber2010classification,
23_wismuller2009exploration,
24_bunte2011neighbor,
25_meyer2004model,
wismuller2001exploration,
saalbach2005hyperbolic,
wismuller2009computational,
26_wismuller2009computational,
leinsinger2003volumetric,
27_meyer2003topographic,
wismuller2009exploration,
28_meyer2009small,
29_wismueller2010model,
meyer2007unsupervised,
wismuller2000neural,
meyer2007analysis,
32_wismueller2008human,
wismuller2000hierarchical,
wismuller2015method,
33_huber2012texture,
37_otto2003model,
38_varini2004breast,
48_meyer2004computer,
40_meyer2004stability,
wismuller1998hierarchical,
41_meyer2008computer,
wismuller2013introducing,
45_bhole20143d,
46_nagarajan2013computer,
wismuller2001automatic,
wismueller1999adaptive,
meyer2005computer,
49_nagarajan2014computer,
pester2013exploring,
50_nagarajan2014classification,
yang2014improving,
wismuller2014pair,
wang2014investigating,
51_wismuller2014framework,
meyer2004local,
schmidt2014impact,
nagarajan2015integrating,
wismuller2015nonlinear,
nagarajan2015characterizing,
abidin2015volumetric,
wismuller2016mutual,
abidin2016investigating,
52_schmidt2016multivariate,
abidin2017classification,
abidin2017using,
61_dsouza2017exploring,
55_abidin2018deep,
chockanathan2018resilient,
dsouza2018mutual,
abidin2019investigating,
abidin2020detecting,
wismuller2020prospective,
dsouza2020large,
vosoughi2020large,
vosoughi2021marijuana, 
vosoughi2021eusipco,
dsouza2021large, 
vosoughi2021_LSAGC,
vosoughi2021_lsNonlinear,
vosoughi2021schizophrenia
}].
\section{METHODOLOGIC CONCEPT }
\subsection{Shortcomings of Current Approaches to Radiology QA}
The potential of conventional approaches to radiology QA is limited and subject to manifold shortcomings. A typical scenario for contemporary QA-related analysis would be to randomly select cases for human expert review, where medical imaging studies are compared with the findings described in the associated radiology reports. Although the scope of the analysis can be restricted to specific imaging study subsets, such as originating from given pre-defined radiology practices, scanners, imaging modalities, radiology exam types, patient origin (inpatient, outpatient, or emergency department), specific referring physicians, etc., the principal challenge remains the same, namely a lack of focus by not pre-selecting cases with a high ‘pre-test’ probability of discordant interpretation based on image content. 
A classic example are ‘peer-review’ approaches, such as the American College of Radiology (ACR) RADPEERTM program [\citeonline{2_SPIE22}], which restricts QA to cases with existing prior imaging studies. This program starts from the assumption that if, during interpretation of a new examination, there are prior images of the same area of interest, the interpreting radiologist will typically form an opinion of the previous interpretation while interpreting the new study. If the opinion of the previous interpretation is scored, a peer review event has occurred, where the reviewer scores the previous interpretation using a standardized rating scale. This approach requires significant human effort, excludes studies with no prior exams, does not distinguish between high and low ‘pre-test’ probability for discordant reads, so that only a tiny subset of inaccurate interpretations may be detected, given the usually low probability of clinically relevant discordant interpretations. In practice, most of the reviewed studies will be rated as ACR category 1 signifying a concordant read. This involves the issue of vigilance fatigue for rare events, e.g., [\citeonline{3_SPIE22_freudenburg1988perceived}], as known from non-medical domains, such as policing [\citeonline{4_SPIE22_krause2012vigilance}] or airport security, where the underlying probability of alert-inducing events is usually low. Case selection is usually based on a random approach, such as automatically requesting peer review for a certain number of cases (typically two) at the beginning of a radiologist’s shift, which will preclude the majority of inaccurate interpretations from being detected in the peer review process. In addition, such a peer-review approach does not only require human intervention by the original peer-reviewing radiologist, but also significant time and effort by a second human reviewer that performs a secondary review of cases with peer-review scores indicating discordant reads between current and prior studies. 
In conclusion, despite requiring a significant amount of human expert intervention time and effort, the vast majority of inaccurate radiology interpretations will remain undetected under the currently used radiology QA approaches. Hence, there is an urgent need to make the radiology QA process better and faster, with the goal to both improve patient safety and reduce costs 
\subsection{Artificial Intelligence (AI)-Based Quality Assurance by Restricted Investigation of Unequal Scores (AQUARIUS)}
AQUARIUS consists of two steps: In a first step, AI-based case analysis is compared to a ‘gold standard’ given by an automatically retrieved estimate of human case assessment. In a second step, cases with “unequal scores”, i.e., discordant assessments between AI-based and human interpretation, are reviewed by human experts. A typical diagnostic radiology implementation of AQUARIUS would include automatic comparison of AI-based image analysis results with natural language processing (NLP) results on final radiology reports. Only the usually small subset of cases with discordant reads would then be manually reviewed by dedicated subspecialty-trained expert radiologists. By restricting human effort to only those cases with a high ‘pre-test’ probability for missed or overcalled findings, AQUARIUS optimizes the use of human and fiscal resources spent on radiology QA programs, thus demonstrating a significant potential for both ‘saving lives’ and ‘saving money’ simultaneously. 
We emphasize that multiple organizational and technical challenges have to be met simultaneously for implementing a successful AQUARIUS study design. These challenges include, but are not limited to, (i) acquiring Institutional Review Board approval for automatic data retreival and radiologists’ participation in such QA initiatives, (ii) aligning the interests of diverse stakeholders, including radiologists, hospital administrators, AI solution vendors, and regulatory bodies, such as the ACR, and (iii) significant interface programming efforts for automatically retrieving heterogeneous information from multiple radiology and hospital IT systems, such as AI-based image analysis results and NLP-based classification of radiology reports for relevant clinical imaging findings.

\section{CLINICAL APPLICATION: INTRACRANIAL HEMORRHAGE DETECTION TRIAL }

\paragraph{Clinical Relevance:} To demonstrate the applicability of our novel hybrid human-AI quality assurance (QA) approach using the AQUARIUS framework, we performed a study for evaluating radiologists' performance on accurately reporting intracranial hemorrhage (ICH) in emergent care setting head CT scans. Timely detection of ICH on medical imaging studies is critical, because delayed therapeutic interventions in emergency settings may be detrimental for patient outcome. For example, in hemorrhagic stroke, the American Heart Association (AHA)/American Stroke Association (ASA) 2018 guidelines for early management emphasize the time-dependent benefit of tissue plasminogen activator (tPA) therapy [\citeonline{5_SPIOE22_powers20182018}]. Patients with ischemic stroke must be identified as soon as possible so that tPA can be administered within 3-4.5 hours of symptom onset [\citeonline{5_SPIOE22_powers20182018}]. A non-contrast head CT scan, which must be negative for hemorrhage, is a required diagnostic test before tPA can be administered in a timely manner [\citeonline{6_SPIE22_lees2010time,7_SPIE22_walter2012diagnosis,8_SPIE22_katzan2004utilization}]. The clinical outcomes of patients receiving tPA within a designated time limit are closely linked to time of treatment [\citeonline{8_SPIE22_katzan2004utilization,9_SPIE22_mazighi2013impact}]. Besides hemorrhagic stroke, there are numerous other clinical conditions, in which ICH will lead to rapid increase of intracranial pressure, where early detection and treatment of ICH can significantly reduce patient morbidity and mortality.  
\paragraph{Data and Methods:} A total of 1936 consecutive non-contrast emergency-setting head CT scans from 2 CT scanners at a large academic hospital were prospectively acquired over 47 consecutive days. Immediately following image acquisition, scans were automatically analyzed for ICH using commercially available software (Aidoc, Tel Aviv, Israel). Cases rated positive for ICH by AI (ICH-AI+) were automatically flagged in radiologists' reading worklists, where flagging was randomly switched off with probability 50\%, see e.g. [\citeonline{10_SPIE22_wismuller2020prospective}]. For ICH-AI+ cases, radiologists' missed ICH detection rates (ratio of number of missed ICH cases and number of all true ICH-AI+ cases) was calculated and compared between flagged and non-flagged ICH-AI+ cases, where images of all ICH-AI+ cases with ICH-negative results detected by natural language processing (NLP) of final radiology reports (ICH-AI+NLP- cases) were re-analyzed by an experienced neuroradiologist to identify true ICH+ cases missed by original radiology readings.

\section{RESULTS}\label{sec:results}
Among all 1936 CT scans, 381 ICH-AI+ cases were found, of which 190 cases were flagged. A total of 29 ICH-AI+NLP- cases were found, where 6 had been reported ICH+ by the radiology report. Of the remaining non-reported 23 ICH-AI+ cases, neuroradiology expert review identified 6 non-reported true ICH+ cases, where 5 cases were non-flagged, and only 1 case was flagged. This yields radiologists' missed ICH detection rates of 0.52\% and 2.5\% for flagged and nonflagged cases, respectively.  
\section{CONCLUSIONS}\label{sec:conclusions}
We present a novel hybrid human-machine interoperability approach, \textbf{A}rtificial Intelligence (AI)-Based \textbf{Qu}ality \textbf{A}ssurance by \textbf{R}estricted \textbf{I}nvestigation of \textbf{U}nequal \textbf{S}cores (AQUARIUS), for re-defining radiology QA, which has the potential of reducing human QA effort by several orders of magnitude over existing approaches, making radiology QA programs better and faster. AQUARIUS consists of two steps: In a first step, AI-based case analysis is compared to a ‘gold standard’ given by an automatically retrieved estimate of human case assessment. In a second step, cases with “unequal scores”, i.e., discordant assessments between AI-based and human interpretation, are reviewed by human experts. In our initial clinical study on detecting QA for detecting Intracranial Hemorrhage (ICH) on emergent noncontrast head CT scans, our results suggest that flagging ICH-AI+ cases on radiologists' worklists may decrease the rate of missed true ICH+ cases, although this was not found statistically significant based on small numbers of discordantly reported cases. Yet, our novel method of human expert review of ICH-AI+NLP- cases successfully revealed true ICH+ cases missed by original radiology reports with minimum human effort, and can therefore provide valuable contributions to radiology QA programs as an efficient hybrid human-AI approach. We conclude that, by combining AI image analysis with NLP-based pre-selection of cases for targeted human expert review, our AQUARIUS approach can efficiently identify missed findings in radiology reports and thus expedite radiology QA programs.

\renewcommand{\baselinestretch}{0.5}
\footnotesize

\acknowledgments % equivalent to \section*{ACKNOWLEDGMENTS}       
 
This research was partially funded by the American College of Radiology (ACR) Innovation Award “AI-PROBE: A Novel Prospective Randomized Clinical Trial Approach for Investigating the Clinical Usefulness of Artificial Intelligence in Radiology” (PI: Axel Wism{\"u}ller) and an Ernest J. Del Monte Institute for Neuroscience Award from the Harry T. Mangurian Jr. Foundation (PI: Axel Wism{\"u}ller). This work was conducted as a Practice Quality Improvement (PQI) project related to American Board of Radiology (ABR) Maintenance of Certificate (MOC) for A.W.
\newpage
% References
\bibliography{report,refs_citations} % bibliography data in report.bib
\bibliographystyle{spiebib} % makes bibtex use spiebib.bst

\end{document}